\documentclass[prd,twocolumn,showpacs,nofootinbib,floatfix]{revtex4}

\usepackage{graphicx}
\usepackage{amsmath,amsfonts}
\usepackage{dcolumn}

\newcommand{\td}[1]{{\rm d}#1} 
\newcommand{\Metric}{g} 
\newcommand{\Snormal}{n} 
\newcommand{\Lapse}{\alpha} 
\newcommand{\CLapse}{\tilde\alpha} 
\newcommand{\Shift}{\beta} 
\newcommand{\SMetric}{\gamma} 
\newcommand{\CMetric}{\tilde\gamma} 
\newcommand{\Ccon}[1]{{\tilde\Gamma^{#1}{}}} 
\newcommand{\CF}{\psi} 
\newcommand{\ExCurv}{K} 
\newcommand{\TrExCurv}{K} 
\newcommand{\TFExCurv}{A} 
\newcommand{\CTFExCurv}{\tilde{A}} 
\newcommand{\SRicci}{\bar{R}} 
\newcommand{\CRicci}{\tilde{R}} 
\newcommand{\SRicciS}{\bar{R}} 
\newcommand{\CRicciS}{\tilde{R}} 
\newcommand{\dtime}{\partial_t} 
\newcommand{\SCD}{{\bar\nabla\!}} 
\newcommand{\CCD}{{\tilde\nabla}\!} 
\newcommand{\LieD}[1]{{{\cal L}_{#1}}}
\newcommand{\CLD}[1]{(\tilde{\mathbb L}{#1})} 

\newcommand{\CMtd}{\tilde{u}} 
\newcommand{\CBnormal}{\tilde{s}} 
\newcommand{\CBMetric}{\tilde{h}} 
\newcommand{\CBExCurv}{\tilde{H}} 
\newcommand{\CBCD}{\tilde{D}} 
\newcommand{\CBcon}[1]{{\hat\Gamma^{#1}{}}} 
 
\newcommand{\FMetric}{f} 
\newcommand{\FCD}{{{}_{\mbox{\tiny f}\!\!}\nabla\!}} 
\newcommand{\Fcon}[1]{{{}_{\mbox{\tiny f}\!}\Gamma^{#1}{}}} 
\newcommand{\Rcon}[1]{{\delta\tilde\Gamma^{#1}{}}} 
\newcommand{\RLap}{{{}_{\mbox{\tiny f}}\Delta}} 
\newcommand{\GaugeFunc}{V} 
\newcommand{\FBMetric}{S} 
\newcommand{\FBCD}{{{}_{\mbox{\tiny s}\!\!}\nabla\!}} 
\newcommand{\FBcon}[1]{{{}_{\mbox{\tiny s}\!}\Gamma^{#1}{}}} 
\newcommand{\RadialLapse}{\Lambda} 
\newcommand{\RadialShift}{\lambda} 

\begin{document}

\title{Excision boundary conditions for the conformal metric}

\author{Gregory B. Cook}\email{cookgb@wfu.edu}
\affiliation{Department of Physics, Wake Forest University,
		 Winston-Salem, North Carolina 27109}

\author{Thomas W. Baumgarte}\email{tbaumgar@bowdoin.edu}
\altaffiliation{Also at Department of Physics, University of Illinois, Urbana, Il 61801}
\affiliation{Department of Physics and Astronomy, Bowdoin College,
Brunswick, Maine 04011}

\date{\today}

\begin{abstract}
Shibata, Ury\=u and Friedman recently suggested a new decomposition of
Einstein's equations that is useful for constructing initial data.  In
contrast to previous decompositions, the conformal metric is no longer
treated as a freely-specifiable variable, but rather is determined as
a solution to the field equations.  The new set of freely-specifiable
variables includes only time-derivatives of metric quantities, which
makes this decomposition very attractive for the construction of
quasiequilibrium solutions.  To date, this new formalism has only been
used for binary neutron stars.  Applications involving black holes
require new boundary conditions for the conformal metric on the domain
boundaries.  In this paper we demonstrate how these boundary
conditions follow naturally from the conformal geometry of the
boundary surfaces and the inherent gauge freedom of the conformal
metric.
\end{abstract}


\maketitle

\section{Introduction}
\label{sec:introduction}

Most numerical relativity applications solve Einstein's field
equations with the help of a 3+1 decomposition, which slices the
spacetime into a foliation of spatial hypersurfaces and splits the
equations into a set of constraint equations and a set of evolution
equations \cite{ADM,york79}.  The constraint equations restrict the
geometry of each hypersurface, representing an instant of constant
coordinate time, while the evolution equations determine how the
geometry changes from one hypersurface to the next.  In its simplest
form, constructing initial data for an evolution calculation therefore
requires finding a solution to the constraint equations.

The Einstein constraints form a set of four equations -- one in the
Hamiltonian constraint and three in the momentum (or vector)
constraint -- and therefore can determine only four of the initial-data
variables; the remaining variables are freely specifiable and have to
be chosen independently before the constraint equations can be solved.
A decomposition of the constraint equations separates the freely
specifiable variables from the constrained ones.  Given a particular
decomposition, the construction of initial data then entails making
well-motivated choices for the freely-specifiable variables that
encode the physical characteristics of the system one wishes to model.

Most decompositions of the constraint equations conformally decompose
the spatial metric (see Sec.~\ref{sec:3+1-conf-decomp} for details)
and treat the conformally related metric as a freely-specifiable
variable, meaning that this ``background'' geometry can be chosen
arbitrarily.  In most applications, the conformally related metric is
then chosen to be flat. This choice simplifies the equations
dramatically, and while it may encode a certain amount of physically
unrealistic gravitational radiation in the initial data -- which often
manifests itself as ``junk radiation'' when the data are evolved -- it
is rarely clear a priori how to make a better choice.

An attractive alternative was recently proposed by Shibata, Ury\=u and
Friedman (\cite{Shibata-etal-2004}; hereafter SUF).  In their new
initial-data decomposition, the conformal metric is no longer treated
as a freely specifiable variable, and is instead determined during the
solution of the equations.  Aside from the trace of the extrinsic
curvature, the new set of freely specifiable variables includes only
time derivatives of the metric and extrinsic curvature.  For the
construction of equilibrium or quasiequilibrium data it is much more
natural to specify the time derivative of metric quantities rather
than the quantities themselves, which makes this new decomposition
very appealing.

So far, this formalism has been used only for binary neutron stars
\cite{Uryu-etal-2006}.  When solving the constraint equations for
black holes, the black hole interior is often excised to avoid
singularities, which requires suitable boundary conditions on the
black hole horizons (see
\cite{Cook-2002,cook-pfeiffer-2004a,Jaramillo-etal:2004,DainJK05,caudill-etal-2006,GouJ06,JarAL07}).
In the context of constrained evolution, excision boundary conditions
for all variables, including the spatial metric, were explored in
Ref.~\cite{Jaramillo-etal:2008}.  Because the spatial metric is
evolved in this case, and the spacelike boundaries used have no
incoming characteristics, no boundary conditions on the spatial metric
were needed.  However, since the new SUF formalism treats the spatial
conformal metric as a constrained variable rather than as either a
freely-specifiable metric or an evolved quantity -- and hence as a
solution of an elliptic equation -- we need to provide suitable
boundary conditions for the conformal metric on black-hole excision
boundaries before this formalism can be applied to black holes.  In
this paper we demonstrate how these boundary conditions can be
formulated quite naturally in terms of the conformal geometry of the
excision boundaries and the inherent gauge freedom in the conformally
related metric.

We found it useful to develop this formalism in terms of a reference
metric approach.  While this approach is not new to general
relativity, it has not been used widely within the numerical
relativity community.  We will therefore present a full description of
the reference metric approach and write all of the Einstein evolution
and constraint equations in terms of this formalism.

Finally, we must keep in mind that for coupled non-linear elliptic
equations, it is not always clear whether or not a given set of
boundary conditions are independent and lead to a well-posed elliptic
system.  In Refs.~\cite{Cook-2002} and \cite{Jaramillo-etal:2004}, two
different physically motivated boundary conditions were developed for
the lapse for use with the extended conformal thin-sandwich equations
(see Sec.~\ref{sec:ID-formulations}).  When
combined with the boundary conditions for the remaining elliptic
variables, however, neither condition gave rise to an independent set of boundary conditions,
and hence left 
the system degenerate (see Refs.~\cite{cook-pfeiffer-2004a} and
\cite{JarAL07}, and also Ref.~\cite{Dain-2006} for a general
discussion of elliptic systems).  While we will derive a
well-motivated set of boundary conditions for the conformal metric, we
have not yet shown that they lead to a well-posed elliptic system.

This article is organized as follows.  We provide an overview of the
problem in Sec.~\ref{sec:overview} and derive the reference metric
approach in Sec.~\ref{sec:ref_metric}.  We then derive the boundary
conditions on the conformally related metric in
Sec.~\ref{sec:boundary-conditions}, and briefly summarize in
Sec.~\ref{sec:summary}.  We also include a complete list of Einstein's
equations in the reference metric approach in App.~\ref{sec:app}.

\section{Overview}
\label{sec:overview}

\subsection{$3+1$ and conformal decompositions}
\label{sec:3+1-conf-decomp}

We begin with the spacetime metric written in the general $3+1$ form
\begin{equation}\label{eq:3+1metric}
\td{s}^2 = -\Lapse^2 \td{t}^2 
  + \SMetric_{ij}(\td{x}^i +\Shift^i\td{t})(\td{x}^j +\Shift^j\td{t}),
\end{equation}
which treats the full spacetime as a foliation of spacelike slices
with timelike unit normal vector $\Snormal^\mu$.  In
Eq.~(\ref{eq:3+1metric}), $\Lapse$ is the scalar lapse of time,
$\Shift^i$ is the shift vector, and $\SMetric_{ij}$ is the metric of a
spatial slice.  In terms of this decomposition, the vacuum Einstein
equations are written as
\begin{eqnarray}
\label{eq:EE_dtK}
\dtime\ExCurv_{ij} &=& -\SCD_i\SCD_j\Lapse 
+ \Lapse\left[\SRicci_{ij} 
- 2\ExCurv_{i\ell}\ExCurv^\ell_j 
+ \TrExCurv\ExCurv_{ij}\right] 
\\ \nonumber && \mbox{}
+\Shift^\ell\SCD_\ell\ExCurv_{ij} 
+ 2\ExCurv_{\ell(i}\SCD_{j)}\Shift^\ell \\
\label{eq:EE_dtg}
\dtime\SMetric_{ij} &=& -2\Lapse\ExCurv_{ij} + 2\SCD_{(i}\Shift_{j)} \\
\label{eq:EE_HC}
0 &=& \SRicciS + \TrExCurv^2 - \ExCurv_{ij}\ExCurv^{ij} \\
\label{eq:EE_MC}
0 &=& \SCD_j(\ExCurv^{ij} - \SMetric^{ij}\TrExCurv)
\end{eqnarray}
Equations (\ref{eq:EE_HC}) and (\ref{eq:EE_MC}) are, respectively, the
Hamiltonian and momentum constraint equations, and
Eq.~(\ref{eq:EE_dtK}) is the evolution equation for the extrinsic
curvature $\ExCurv_{ij}$ which is itself defined by
Eq.~(\ref{eq:EE_dtg}).  The covariant derivative compatible with the
spatial metric $\SMetric_{ij}$ is written as $\SCD_i$, the Ricci
tensor and scalar curvature as $\SRicci_{ij}$ and $\SRicciS$, and
$\dtime$ is the derivative along the time vector
\begin{equation}
t^\mu = \Lapse \Snormal^\mu + \Shift^\mu.
\end{equation}
Finally, it is often convenient to decompose the extrinsic curvature
into its trace $\TrExCurv\equiv\SMetric^{ij}\ExCurv_{ij}$ and trace-free
parts
\begin{equation}
\TFExCurv_{ij} \equiv \ExCurv_{ij} - \mbox{$\frac13$}\SMetric_{ij}\TrExCurv.
\end{equation}

It is also useful to conformally decompose the spatial metric
and other quantities.  In particular, we will make use of the
conformal factor $\CF$ which allows us to define the
conformal metric $\CMetric_{ij}$, conformal trace-free extrinsic
curvature $\CTFExCurv_{ij}$ and conformal lapse $\CLapse$ via
\begin{eqnarray}
\SMetric_{ij} &\equiv& \CF^4\CMetric_{ij} \\
\TFExCurv_{ij} &\equiv& \CF^{-2}\CTFExCurv_{ij} \\
\Lapse &\equiv& \CF^6\CLapse.
\end{eqnarray}

In terms of these conformal variables and the covariant derivative
$\CCD_i$, Ricci tensor $\CRicci_{ij}$ and scalar curvature $\CRicciS$ 
compatible with the conformal metric $\CMetric_{ij}$, we
can write the various constraint and evolution equations as
\begin{eqnarray}
\label{eq:CHamCon1}
0 &=& \CCD\,^2\CF - \mbox{$\frac18$}\CF\CRicciS
- \mbox{$\frac1{12}$}\CF^5\TrExCurv^2
+ \mbox{$\frac18$}\CF^{-7}\CTFExCurv_{ij}\CTFExCurv^{ij} \\
\label{eq:CMomCon1}
0 &=& \CCD_j\CTFExCurv^{ij} 
     - \mbox{$\frac23$}\CF^6\CMetric^{ij}\CCD_j\TrExCurv \\
\label{eq:dtCmet1}
\dtime\CMetric_{ij} &=& -2\CLapse\CTFExCurv_{ij} 
+ \CLD{[\CF^{-4}\Shift]}_{ij} \\
\label{eq:dtCA1}
\dtime\CTFExCurv^{ij} &=& 
\CF^8\CLapse(\CRicci^{ij} - \mbox{$\frac13$}\CMetric^{ij}\CRicciS)
\\ && \mbox{}
- \left[\CCD\,^i\CCD\,^j
- \mbox{$\frac13$}\CMetric^{ij}\CCD\,^2\right](\CF^8\CLapse) 
\nonumber \\ && \mbox{}
+ 8\CF^8\CLapse\left[(\CCD\,^{(i}\ln\CF)\CCD\,^{j)}\ln(\CF^7\CLapse)
\right.\nonumber \\ && \mbox{} \hspace{0.75in}\left.
- \mbox{$\frac13$}\CMetric^{ij}(\CCD\,^k\ln\CF)\CCD_k\ln(\CF^7\CLapse)\right] 
\nonumber \\ && \mbox{}
+ \Shift^k\CCD_k\CTFExCurv^{ij} + \CTFExCurv^{ij}\CCD_k\Shift^k
+ \CTFExCurv_k^{(i}\CCD\,^{j)}\Shift^k
\nonumber \\ && \mbox{}
- \CTFExCurv^{k(i}\CCD_k\Shift^{j)}
 + \CTFExCurv_k^{(i}\dtime\CMetric^{j)k} 
\nonumber \\
\label{eq:dtTrK1}
\dtime\TrExCurv &=& -\CF^{-5}\CCD\,^2(\CF^7\CLapse)
+ \mbox{$\frac18$}\CF^2\CLapse\CRicciS
+ \mbox{$\frac5{12}$}\CF^6\CLapse\TrExCurv^2 
\\ && \mbox{}
+ \mbox{$\frac78$}\CF^{-6}\CLapse\CTFExCurv_{ij}\CTFExCurv^{ij}
+ \Shift^k\CCD_k\TrExCurv \nonumber \\
\dtime\CF &=& \mbox{$\frac16$}\CF(\CCD_k\Shift^k 
+ 6\Shift^k\CCD_k\ln\CF- \CF^6\CLapse\TrExCurv).
\end{eqnarray}
Here the conformal longitudinal derivative of a 1-form $\CLD{\ }_{ij}$ is defined by
\begin{equation}
  \CLD{V}_{ij} \equiv 2\CCD_{(i}V_{j)}
      -\mbox{$\frac23$}\CMetric_{ij}\CMetric^{k\ell}\CCD_kV_\ell,
\end{equation}
and we note that the term $\CF^{-4}$ appears in its argument in Eq.~(\ref{eq:dtCmet1}) because the index of the shift $\Shift^i$ has to be lowered with the physical spatial metric $\SMetric_{ij}$ rather than with the conformal metric $\CMetric_{ij}$.

\subsection{Current initial data formulations}
\label{sec:ID-formulations}

The most commonly used decompositions of the constraint equations are
the conformal transverse-traceless and the conformal thin-sandwich
decompositions (see, e.g., \cite{cook-2000,BaumgarteShapiro:2003,Pfe05} for
reviews).

The conformal transverse-traceless decomposition solves the
Hamiltonian and momentum constraints, Eqs.~(\ref{eq:CHamCon1}) and
(\ref{eq:CMomCon1}), and fixes the conformal metric $\CMetric_{ij}$ as
well as the mean curvature $\TrExCurv$ and the transverse-traceless
parts of $\CTFExCurv_{ij}$ as the freely-specifiable variables.  If
the conformal metric is chosen to be flat, $\CMetric_{ij} = f_{ij}$,
and the initial hypersurface to be maximal, so that $\TrExCurv=0$, the
momentum constraints decouple from the Hamiltonian constraint and
become linear.  Vacuum solutions describing one or multiple black
holes, known as Bowen-York solutions \cite{bowenyork80}, can then be
found analytically and form the basis for both
puncture \cite{brandt_bruegmann97,BeiO94,BeiO96} and conformal-imaging
solutions \cite{bowenyork80,cook91} to the Hamiltonian constraint.

In the conformal thin-sandwich decomposition \cite{york-1999} the
evolution equation for the conformal metric, Eq.~(\ref{eq:dtCmet1}),
is used to replace the transverse-traceless parts of $\CTFExCurv_{ij}$
with the time derivative of the conformally related metric in the set
of freely-specifiable variables.  Effectively, we can solve
Eq.~(\ref{eq:dtCmet1}) for $\CTFExCurv_{ij}$ and then insert this
equation into every occurrence of $\CTFExCurv_{ij}$ in the other
equations.  We now fix the trace of the extrinsic curvature
$\TrExCurv$, the conformal metric $\CMetric_{ij}$, its time derivative
$\dtime\CMetric_{ij}$, as well as the conformal lapse $\CLapse$ as the
freely specifiable data and solve the Hamiltonian constraint
(\ref{eq:CHamCon1}) for the conformal factor $\CF$ and the momentum
constraint (\ref{eq:CMomCon1}), via (\ref{eq:dtCmet1}), for the shift
$\Shift^i$.

The {\em extended} conformal thin-sandwich
decomposition \cite{Pfeiffer-York-2003} incorporates the time
derivative of the trace of the extrinsic curvature,
Eq.~(\ref{eq:dtTrK1}).  If we take $\dtime\TrExCurv$ as specified,
this equation yields an elliptic equation for the conformal lapse.
Essentially, we now replace the conformal lapse with $\dtime\TrExCurv$
in the subset of the freely specifiable variables, and we obtain one
additional elliptic equation which must now be solved for the
conformal lapse. While this change of perspective results in an
additional elliptic equation which must be solved in order to
construct initial data, this added burden is offset by the fact that
the set of freely specifiable data now takes on a particularly
attractive form.  Specifically, we now specify the conformal metric
and its time derivative, along with the trace of the extrinsic
curvature and its time derivative.  In many situations, in particular
as mentioned before for the construction of equilibrium or
quasiequilibrium initial data, it is advantageous to specify the time
derivative of some quantity rather than some other field.  (See
\cite{PfeY05,BauOP07,Wal07,CorCDJNG08} for a discussion of the uniqueness
issue in the extended conformal thin-sandwich formalism.)

\subsection{Determining the conformal metric}

In all of the initial-data methods sketched out in
Sec.~\ref{sec:ID-formulations}, the conformal metric is taken as part
of the subset of freely-specifiable initial data.  Roughly speaking,
the five degrees of freedom that are fixed by specifying the conformal
metric include the initial choice of the spatial gauge and two
dynamical degrees of freedom.  Therefore, fixing the conformal metric
strongly affects the initial gravitational radiation content of the
initial data.

In most cases, the conformal metric is chosen to be flat.  But even
when it is not chosen to be flat, in all but the most trivial cases,
the chosen metric is not completely compatible with the physics that
one wishes to build into the initial data.  The result is that
undesired ``junk'' gravitational radiation is built into the initial
data.  While this defect in the data is usually small and has little
effect on the gross physics one wishes to simulate, it can have a
significant effect on detailed comparisons with post-Newtonian
methods \cite{Boyle-etal-2007}, and may impact future
parameter estimation efforts.

It is desirable to find a way to construct initial data in which the
conformal metric is not fixed {\em a priori}, but is constructed in a
way that is consistent with the physics one wishes to simulate and
that eliminates, or at least reduces, the junk radiation.  Following
in the spirit of the extended conformal-thin-sandwich approach, SUF
noticed that the equation for the time derivative of the trace-free
part of the extrinsic curvature, Eq.~(\ref{eq:dtCA1}), can be written
as an elliptic equation for the conformal metric if the spatial gauge
is imposed in a suitable way and if we take $\dtime\CTFExCurv^{ij}$ as
freely-specified data.  In close analogy to going from the conformal
transverse-traceless to the extended conformal thin-sandwich
decomposition, namely using the evolution equation for the conformal
metric to replace $\CTFExCurv_{ij}$ with $\dtime\CMetric_{ij}$, and
the evolution equation for the trace of the extrinsic curvature to
replace $\CLapse$ with $\dtime\TrExCurv$ as freely specifiable
variables, we now use the evolution equation for the extrinsic
curvature to replace $\CMetric_{ij}$ with $\dtime\CTFExCurv_{ij}$.
This formalism is very attractive since it is again more natural to
make choices for time derivatives of functions than for the functions
themselves.

To date, this approach for constructing initial data has been
implemented successfully only for binary neutron
stars \cite{Uryu-etal-2006}.
The main goal of this paper is to develop the formalism necessary for applications 
to black-hole initial data where the black-hole
interiors are excised from the computational domain.  When black-hole
initial data are computed using excision methods, boundary conditions
for all of the initial data that are determined by elliptic equations
must be applied on the excision boundaries.  For the case of the
extended conformal-thin-sandwich initial data, the boundary conditions
have been worked out and thoroughly tested
\cite{Cook-2002,cook-pfeiffer-2004a,caudill-etal-2006,Jaramillo-etal:2004}.
In the context of the new formalism of SUF, however, additional
boundary conditions are required for the conformal metric.

\section{Reference Metric Approach}
\label{sec:ref_metric}

Our goal is to derive a system of equations that will determine the
conformal metric together with some other quantities.  Many of the
operators that need to be inverted to construct the conformal metric
depend on the conformal metric themselves.  It is therefore convenient
(though certainly not necessary) to use a reference metric approach
wherein, in addition to the conformal metric $\CMetric_{ij}$, we also
associate some appropriate fixed metric with our solution manifold.
We can then formulate the operators in terms of this fixed metric,
which greatly simplifies the inversion of the operators.  This
approach is not new (cf
Refs.~\cite{papadopoulos-Sopuerta-2002,andersson-moncrief-2003}), but
is not widely used within the numerical relativity community.  A notable exception is 
\cite{BGGN2004}, whose formalism shares many elements with ours.  We will
give a basic outline of the approach in this Section, and list the
complete set of Einstein's equations in Appendix \ref{sec:app}.

\subsection{Basic outline}

We assume that our initial data hypersurface is represented at a basic
level by a manifold with coordinates and coordinate maps defined
everywhere.  We associate with this manifold two metrics which are not
necessarily the same: $\Metric_{ij}$ and $\hat\Metric_{ij}$.  Each
metric has an inverse and covariant derivative such that
\begin{eqnarray}
  \Metric_{jk}\Metric^{ik}=\delta^i_j \quad&\mbox{and}&\quad
  \hat\Metric_{jk}\hat\Metric^{ik}=\delta^i_j \\
  \nabla_k\Metric_{ij}=0 \quad&\mbox{and}&\quad \hat\nabla_k\hat\Metric_{ij}=0.
\end{eqnarray}
The difference between two connections is a tensor which can be written
as
\begin{eqnarray}
  \delta\Gamma^k{}_{ij}&\equiv&\Gamma^k{}_{ij}-\hat\Gamma^k{}_{ij}\\ \nonumber
       &=& \frac12\hat\Metric^{k\ell}\left[
             \hat\nabla_i\Metric_{j\ell} + \hat\nabla_j\Metric_{i\ell}
	     - \hat\nabla_\ell\Metric_{ij}\right].
\end{eqnarray}
The differences between the Riemann and Ricci tensors for each metric
can be written as
\begin{eqnarray}
{\delta R_{ijk}}^\ell &\equiv& {R_{ijk}}^\ell-{\hat{R}_{ijk}}{}^\ell 
\\ \nonumber &=&
\hat\nabla_j{\delta\Gamma^\ell}_{ki} - \hat\nabla_i{\delta\Gamma^\ell}_{kj}
\\ \nonumber && \mbox{} \hspace{0.5in}
     + {\delta\Gamma^m}_{ki}{\delta\Gamma^\ell}_{mj} 
     - {\delta\Gamma^m}_{kj}{\delta\Gamma^\ell}_{mi} \\
{\delta R}_{ij} &\equiv& {R_{i\ell j}}^\ell-{\hat{R}_{i\ell j}}{}^\ell
\\ \nonumber &=&
\hat\nabla_\ell{\delta\Gamma^\ell}_{ji} 
     - \hat\nabla_i{\delta\Gamma^\ell}_{j\ell}
\\ \nonumber && \mbox{} \hspace{0.5in}
     + {\delta\Gamma^m}_{ji}{\delta\Gamma^\ell}_{m\ell} 
     - {\delta\Gamma^m}_{j\ell}{\delta\Gamma^\ell}_{mi}.
\end{eqnarray}
The latter can be rewritten as
\begin{eqnarray}
\label{eq:ricci-harmonic}
{\delta R}_{ij}\!\! &=& \!\! - \hat{R}_{ij} 
  - \Metric^{\ell m}\Metric_{k(i} {\hat{R}_{j)\ell m}}{}^k
  - \frac12\Metric^{\ell m}\hat\nabla_\ell\hat\nabla_m \Metric_{ij}
\\ \nonumber &&\!\! \mbox{}
  + \hat\nabla_{(i}\left( \Metric_{j)k} \Metric^{\ell m} {\delta\Gamma^k}_{\ell m}\right)
  - \Metric_{pk} \Metric^{\ell m} {\delta\Gamma^k}_{\ell m}{\delta\Gamma^p}_{ij}
\\ \nonumber &&\!\! \mbox{}
  + \frac12 \Metric^{\ell m}\Metric^{np}\Bigl\{
          (\hat\nabla_i \Metric_{mp})\hat\nabla_\ell \Metric_{jn}
        + (\hat\nabla_j \Metric_{mp})\hat\nabla_\ell \Metric_{in}
\\ \nonumber &&\!\! \mbox{}\hspace{0.75in}
	+ (\hat\nabla_\ell \Metric_{in})\hat\nabla_m \Metric_{jp}
	- (\hat\nabla_\ell \Metric_{in})\hat\nabla_p \Metric_{jm}
\\ \nonumber &&\!\! \mbox{}\hspace{0.75in}
	- \frac12(\hat\nabla_i\Metric_{\ell n})\hat\nabla_j\Metric_{mp}\Bigr\}.
\end{eqnarray}
As it turns out, this form is particularly useful because all second
derivatives of the $\Metric_{ij}$ have now been absorbed into only two
terms.  The first of these two terms, $\Metric^{\ell
m}\hat\nabla_\ell\hat\nabla_m \Metric_{ij}$, forms an elliptic
operator acting on $\Metric_{ij}$ as long as both $\Metric_{ij}$ and
$\hat\Metric_{ij}$ are sufficiently well behaved, while the second of
the two terms, $\hat\nabla_{(i}\left( \Metric_{j)k} \Metric^{\ell m}
{\delta\Gamma^k}_{\ell m}\right)$, can be eliminated by virtue of a
suitable gauge choice, as we will discuss in the next section.

\subsection{Reference metric and gauge choice}

We will assume that we are dealing with one or more black holes in an
asymptotically flat initial-data hypersurface.  In this case, the
solution domain is topologically $E^3$ with a ``ball'' cut out for
each excised black hole interior.  It is therefore appropriate to take
the reference metric $\hat\Metric_{ij}$ to be a flat metric
$\FMetric_{ij}$ (not necessarily in Cartesian coordinate).  For
consistency with our previous notation, the ``other'' metric
$\Metric_{ij}$ will be the conformal metric we wish to determine,
$\CMetric_{ij}$.  To make the notation as clear as possible, we will
denote the covariant derivative compatible with the flat reference
metric as $\FCD_k$, and so the difference of connections becomes
\begin{eqnarray}
\Rcon{k}_{ij} &\equiv& \Ccon{k}_{ij} - \Fcon{k}_{ij} \\ \nonumber
  &=& \mbox{$\frac12$}\CMetric^{k\ell}\left[
\FCD_i\CMetric_{j\ell} + \FCD_j\CMetric_{i\ell} - \FCD_\ell\CMetric_{ij}
\right].
\end{eqnarray}

We next impose a spatial gauge condition by setting the contractions
of the connection coefficients equal to some predetermined gauge
source functions $\GaugeFunc^k$,
\begin{equation}
\label{eq:genconfharm}
\Rcon{k} \equiv \CMetric^{ij}\Rcon{k}_{ij} 
           = - \frac1{\sqrt{\det\CMetric}}
               \FCD_\ell(\sqrt{\det\CMetric}\,\CMetric^{k\ell}) = \GaugeFunc^k.
\end{equation}
For convenience and ease of reading, we will denote the determinant
of the conformal metric as either $\det\CMetric$ or $\CMetric$. One
possible choice is $\GaugeFunc^k = 0$, in which case we would obtain a
``spatial conformal harmonic gauge", but we will leave $\GaugeFunc^k$
arbitrary for generality (compare also the ``generalized Dirac gauge" of 
\cite{BGGN2004}).  With the gauge source functions
$\GaugeFunc^k$ given as specified functions of the coordinates, we see
that the only remaining second-order term in
Eq.~(\ref{eq:ricci-harmonic}) is the first one, bringing the conformal
Ricci tensor into a familiar elliptic form.  This or similar
properties of the Ricci tensor have often been utilized before, both
in a four-dimensional
(e.g.~\cite{Ded21,Lan22,hyper52,fischmars72,friedrich85,Gar02,Pretorius-2005})
and a three-dimensional context
(e.g.~\cite{ShiN95,baumgarte_etal98,andersson-moncrief-2003,BGGN2004}).
An attractive feature of the reference metric approach is the fact
that differences of connection coefficients $\Rcon{k}_{ij}$ are tensors, so that
the gauge source functions $\GaugeFunc^k$ become vectors and take a
gauge-invariant meaning.

In our case, the conformal Ricci tensor and conformal Ricci
scalar take the form
\begin{eqnarray}
\CRicci^{ij}\!\! &=&\!\! \mbox{$\frac12$}\RLap\CMetric^{ij}
+ \CCD\,^{(i}\GaugeFunc^{j)} \\ \nonumber && \mbox{}\hspace{0.4in}
- \mbox{$\frac12$}{\cal B}^{ij} - \mbox{$\frac12$}{\cal C}^{ij} 
- \mbox{$\frac14$}{\cal D}^{ij}
+ {\cal E}^{ij}  \\
\CRicciS\!\! &=&\!\! -\mbox{$\frac12$}\RLap\ln(\det\CMetric)
+ \CCD_k\GaugeFunc^k \\ \nonumber && \mbox{}\hspace{0.4in}
+ \mbox{$\frac12$}\CMetric_{k\ell}\left[{\cal C}^{k\ell}
    - \mbox{$\frac12$}{\cal B}^{k\ell}\right]
\end{eqnarray}
(compare Eqs.~(44) -- (47) in \cite{BGGN2004}).
Here $\RLap$ is a generalized Laplacian defined by
\begin{equation}
\RLap \equiv \CMetric^{k\ell}\FCD_k\FCD_\ell,
\end{equation}
and the symmetric tensors ${\cal B}^{ij}$, ${\cal C}^{ij}$, ${\cal D}^{ij}$, 
and ${\cal E}^{ij}$ are quadratic combinations of first-derivatives
of the conformal metric defined by
\begin{eqnarray}
{\cal B}^{ij} &\equiv& \CMetric_{mn}\CMetric^{k\ell}(\FCD_k\CMetric^{im})
\FCD_\ell\CMetric^{jn} = {\cal B}^{ji} \\
{\cal C}^{ij} &\equiv& (\FCD_k\CMetric^{i\ell})\FCD_\ell\CMetric^{jk}
= {\cal C}^{ji} \\
{\cal D}^{ij} &\equiv& \CMetric^{ip}\CMetric^{jq}
                       \CMetric_{mn}\CMetric_{k\ell}(\FCD_p\CMetric^{mk})
                       \FCD_q\CMetric^{n\ell} = {\cal D}^{ji} \\
{\cal E}^{ij} &\equiv& \CMetric_{mn}(\FCD_k\CMetric^{m(i})
                       \CMetric^{j)\ell}\FCD_\ell\CMetric^{kn}={\cal E}^{ji}.
\end{eqnarray}
We note that the derivatives of
the gauge source functions have been written in terms of the conformal
covariant derivative $\CCD_k$, and not the flat covariant derivative
$\FCD_k$.  Also, in the future, we plan to solve directly for the {\em inverse}
conformal metric $\CMetric^{ij}$ and so the conformal Ricci tensor and
the various derivatives of the metric are written with the indices raised (compare \cite{BGGN2004}).
We list the complete set of Einstein's equations in the reference metric
form in Appendix \ref{sec:app}.

\section{Boundary Conditions}
\label{sec:boundary-conditions}

Our goal in this section is to determine the proper boundary
conditions for the inverse conformal metric $\CMetric^{ij}$ required
on the excision boundaries when we solve an elliptic equation for
this variable.  We begin with some simple counting arguments.

The conformal metric $\CMetric_{ij}$ is a member of a conformal
equivalence class of metrics which each have five independent degrees
of freedom.  To fix upon a particular member of this class, we can
fix the determinant of the conformal metric $\det(\CMetric)$.  As
mentioned previously, these five degrees of freedom are usually
thought of as containing the two dynamical degrees of freedom and the
three spatial gauge degrees of freedom in the metric.  Of course, this
splitting is not clean and, in general, they mix with the sixth
constrained (or longitudinal) degree of freedom of the full spatial
metric $\SMetric_{ij}$.  However, this rough splitting suggests how we
should fix the boundary conditions.

First, the excision surface is a closed 2-surface with $S^2$ topology.
The metric on any such surface 
is conformally equivalent to the unit sphere.  Since the metric
induced on the excision surface by the physical metric $\SMetric_{ij}$
will be conformally equivalent to a spherical metric, it seems that we
should be able to demand that the metric induced on the excision
surface by the conformal metric $\CMetric_{ij}$ be a spherical metric.
In fact, this can be generalized to the statement that we have the
freedom to fix the metric induced on the excision surface by the
conformal metric $\CMetric_{ij}$ to that of any metric with $S^2$
topology.  A 2-metric has three degrees of freedom, but because of
the conformal equivalence of all 2-metrics, specifying the induced
metric really only fixes two degrees of freedom.

Having fixed two of the five degrees of freedom of $\CMetric_{ij}$ on
the excision surface leaves three more boundary conditions that must
be specified.  Since the conformal metric contains three gauge degrees
of freedom, it seems natural that these three remaining boundary
conditions should result from the gauge
conditions~(\ref{eq:genconfharm}).

\subsection{$2+1$ decomposition}

To rigorously define and clearly understand the new boundary
conditions, it is most convenient to rewrite the conformal metric in
terms of a $2+1$ decomposition adapted to the excision surface.  We
define on the solution manifold a foliation of concentric surfaces
with topology $S^2$ in the neighborhood of the excision boundary.
Each surface is defined as a level surface of a scalar function
$r(x^i)$ such that $r=r_0$ defines the excision surface.
Independently of which metric we are using, we have a normal
1-form for the excision surface given by
\begin{equation}
\left.\partial_ir(x^j)\right|_{r_0}.
\end{equation}
In terms of each metric ($\CMetric_{ij}$ and $\FMetric_{ij}$) we can
normalize this 1-form and define a unit-normal vector and unit-normal
1-form.  In terms of the unit-normals to the excision surface, we can
then define projection operators associated with each metric and a
corresponding induced metric on the surface defined by $r=r_0$.  For the
conformal metric $\CMetric_{ij}$, we will denote its induced metric on
the excision surface as $\CBMetric_{AB}$, and for the flat metric
$\FMetric_{ij}$ we will denote the induced metric as $\FBMetric_{AB}$.
Here and in the following, upper-case Latin indices denote adapted
coordinates in the excision surface.

\subsubsection{Conformal metric}
\label{sec:2+1-conformal-metric}

Focusing first on the $2+1$ decomposition of the conformal metric
we write the three-dimensional line interval as
\begin{equation}
   \tilde{\td{s}}^2=\RadialLapse^2\td{r}^2
     + \CBMetric_{AB}(\td{x}^A + \RadialShift^A\td{r})
                     (\td{x}^B + \RadialShift^B\td{r}),
\end{equation}
and note that the determinant is given by
\begin{equation}
\label{eq:2+1-detCMet}
  \det\CMetric =\RadialLapse^2\det\CBMetric.
\end{equation}
Evidently, $\RadialLapse$ and $\RadialShift^A$ play the role of the
lapse $\Lapse$ and shift $\Shift^i$ in the more familiar 3+1
decompositions.  In ``matrix'' form we write the conformal metric as
\begin{equation}
  \CMetric_{ij} = \left[\begin{array}{cc}
      \RadialLapse^2 + \RadialShift_C\RadialShift^C & \RadialShift_B \\
      \RadialShift_A  &  \CBMetric_{AB} \end{array}\right], 
\end{equation}
where $\RadialShift_A\equiv\CBMetric_{AB}\RadialShift^B$, and its inverse as 
\begin{equation}
  \CMetric^{ij} = \left[\begin{array}{cc}
      \RadialLapse^{-2} & -\RadialLapse^{-2}\RadialShift^B \\
      -\RadialLapse^{-2}\RadialShift^A  &  
      \CBMetric^{AB}+\RadialLapse^{-2}\RadialShift^A\RadialShift^B 
    \end{array}\right].
\end{equation}
The unit-normal 1-form is given by 
\begin{equation}
  \CBnormal_i \equiv \RadialLapse\left.\partial_ir(x^j)\right|_{r_0} 
  = \mbox{} \RadialLapse[1,\vec{0}],
\end{equation}
and the unit-normal vector by
\begin{equation}
  \CBnormal^i = \CMetric^{ij}\CBnormal_j 
              = \RadialLapse^{-1}[1,-\RadialShift^A].
\end{equation}
Similarly, the ``radial'' vector is given by
\begin{equation}
  r^i = \RadialLapse\CBnormal^i+\RadialShift^i 
  = \mbox{} [1,\vec{0}],
\end{equation}
where the 3-vector $\RadialShift^i\equiv[0,\RadialShift^A]$.

The induced metric on the excision surface is given by 
\begin{equation}
  \CBMetric_{ij} = \CMetric_{ij} - \CBnormal_i\CBnormal_j,
\end{equation}
and we define the extrinsic curvature of the excision surface as
\begin{equation}
\label{eq:SurfExCurv_def}
   \CBExCurv_{ij}\equiv\CBMetric^k_i\CBMetric^\ell_j\CCD_{(k}\CBnormal_{\ell)}.
\end{equation}
The covariant derivative compatible with the induced metric $\CBMetric_{AB}$
will be denoted $\CBCD_C$.  Because we are using adapted coordinates,
the Lie derivative in the radial direction $\LieD{\vec{r}}$ is equivalent to
the partial derivative in the radial direction,
\begin{equation}
   \LieD{\vec{r}}\equiv \frac{\partial}{\partial r} = r^i\partial_i.
\end{equation}
Finally, we note that
\begin{equation}
\label{eq:SurfExCurv}
   \partial_r\CBMetric_{AB} = 2\RadialLapse\CBExCurv_{AB} 
      + \CBCD_A\RadialShift_B + \CBCD_B\RadialShift_A,
\end{equation}
which follows from the definition of the extrinsic curvature.

Since our gauge conditions (\ref{eq:genconfharm}) are defined in terms
of the difference between the conformal and flat connections, it is useful
to list the connections associated with the conformal metric in terms of
the $2+1$ variables.
\begin{subequations}\label{eq:ConfConns}
\begin{eqnarray}
  \Ccon{r}_{rr} &=& \partial_r\ln\RadialLapse 
     + \RadialShift^A\CBCD_A\ln\RadialLapse
     - \frac1\RadialLapse \RadialShift^A \RadialShift^B\CBExCurv_{AB}, \\
  \Ccon{r}_{rA} &=& \CBCD_A\ln\RadialLapse 
     - \frac1\RadialLapse\RadialShift^B\CBExCurv_{AB}, \\
\label{eq:ConfConnrAB}
  \Ccon{r}_{AB} &=& -\frac1\RadialLapse\CBExCurv_{AB}, \\
  \Ccon{A}_{rr} &=& \partial_r\RadialShift^A 
     - \RadialLapse^2\CBCD^A\ln\RadialLapse 
     + 2\RadialLapse\RadialShift^B\CBExCurv^A_B 
\\ \nonumber && \mbox{} \hspace{0.2in} 
     + \RadialShift^B\CBCD_B\RadialShift^A 
     - \RadialShift^A \Ccon{r}_{rr}, \\
\label{eq:ConfConnArB}
  \Ccon{A}_{rB} &=& -\RadialShift^A \Ccon{r}_{rB} 
     + \RadialLapse\CBExCurv^A_B + \CBMetric^{AC}\CBCD_B\RadialShift_C, \\
  \Ccon{A}_{BC} &=& -\RadialShift^A \Ccon{r}_{BC} + \CBcon{A}_{BC},
\end{eqnarray}
\end{subequations}
where $\CBcon{A}_{BC}$ is the usual connection constructed from the induced
conformal metric $\CBMetric_{AB}$.

\subsubsection{Flat metric}
The $2+1$ decomposition of the flat metric $\FMetric_{ij}$ follows
precisely the same form as that for the conformal metric as outlined
above in Sec.~\ref{sec:2+1-conformal-metric}.  We simply need to define
new variables for the analogs of $\RadialLapse$, $\RadialShift^A$, and
$\CBMetric_{AB}$.  While it is possible to proceed in full generality,
we will restrict ourselves to the case where we take the flat metric
analog of $\RadialLapse\to1$ and $\RadialShift^A\to0$ so that we have
\begin{equation}
  \FMetric_{ij} = \left[\begin{array}{cc}
      1 & 0 \\
      0  &  \FBMetric_{AB} \end{array}\right]
\end{equation}
and
\begin{equation}
  \FMetric^{ij} = \left[\begin{array}{cc}
     1 & 0 \\
      0  & \FBMetric^{AB} \end{array}\right].
\end{equation}
Specifying the induced metric $\FBMetric_{AB}$ is sufficient to fix
the coordinate transformations between the coordinates on the
3-dimensional initial-data manifold and the $2+1$ adapted coordinates.
For example, the simplest choice is to let $\FBMetric_{AB}$ be the
standard metric for a sphere of radius $r$.  Then, if the initial-data
manifold uses standard Cartesian coordinates, the coordinate
transformations are well known.

To proceed, we will need to define the covariant derivative compatible with
$\FBMetric_{AB}$.  We will denote this derivative as $\FBCD_A$.  Finally,
we can write the connections associated with our flat metric as
\begin{subequations}\label{eq:FlatConns}
\begin{eqnarray}
  \Fcon{r}_{rr} &=& 0, \\
  \Fcon{r}_{rA} &=& 0, \\
\label{eq:FlatConnrAB}
  \Fcon{r}_{AB} &=& -\frac12\partial_r\FBMetric_{AB}, \\
  \Fcon{A}_{rr} &=& 0, \\
\label{eq:FlatConnArB}
  \Fcon{A}_{rB} &=& \frac12\FBMetric^{AC}\partial_r\FBMetric_{BC}, \\
  \Fcon{A}_{BC} &=& \FBcon{A}_{BC},
\end{eqnarray}
\end{subequations}
where $\FBcon{A}_{BC}$ is the usual connection constructed from the
induced flat metric $\FBMetric_{AB}$.  Note that
Eqs.~(\ref{eq:FlatConnrAB}) and (\ref{eq:FlatConnArB}) follow from
Eqs.~(\ref{eq:ConfConnrAB}) and (\ref{eq:ConfConnArB}) by using
Eq.~(\ref{eq:SurfExCurv}).

\subsection{Gauge boundary conditions}
All the pieces are now in place to fully understand and rigorously
define excision boundary conditions that can be used for the conformal
metric.  Recall from our discussion at the beginning of
Sec.~\ref{sec:boundary-conditions} that to fix upon a unique member from
the conformal equivalence class of conformal metrics $\CMetric_{ij}$, we
may specify its determinant $\det(\CMetric)$ everywhere in the initial-data
manifold.  We choose to demand that the determinants of the conformal
metric and the reference metric are equal.  In this paper, we are
restricting ourselves to the case that the reference metric is flat,
so we demand
\begin{equation}
\label{eq:ECrestriction}
  \det\CMetric = \det\FMetric.
\end{equation}
We also point the reader to Sec.~II.B of Ref.~\cite{BGGN2004} for
a more in-depth discussion of this point.

Also recall that $\CBMetric_{AB}$ and $\FBMetric_{AB}$, each having
topology $S^2$, are both related to the standard metric of a
unit-sphere by a conformal transformation and an appropriate
coordinate transformation.  Now, since we are expressing both
$\CBMetric_{AB}$ and $\FBMetric_{AB}$ in the same coordinates, we can
demand that they are related to each other simply by a conformal
transformation:
\begin{equation}
\label{eq:Bconfrelat}
  \CBMetric_{AB} = \Omega^2(x^i)\FBMetric_{AB}.
\end{equation}
It follows immediately from Eq.~(\ref{eq:2+1-detCMet}), its analog for
the induced flat metric that $\det(\FMetric)=\det(\FBMetric)$, and
Eq.~(\ref{eq:Bconfrelat}) that
\begin{equation}
\label{eq:BConflapserelat}
  \Omega^2 = \frac1\RadialLapse.
\end{equation}

Notice that because of Eqs.~(\ref{eq:Bconfrelat}) and
(\ref{eq:BConflapserelat}), the $2+1$ version of the conformal metric
can be expressed in terms of $\RadialLapse$, $\RadialShift^A$, and
$\FBMetric_{AB}$:
\begin{equation} \label{eq:ind_metric}
   \tilde{\td{s}}^2=\RadialLapse^2\td{r}^2
     + \frac1{\RadialLapse}\FBMetric_{AB}(\td{x}^A + \RadialShift^A\td{r})
                                           (\td{x}^B + \RadialShift^B\td{r}).
\end{equation}
These variables hold six degrees of freedom.  But
recall that they incorporate the restriction on the conformal
equivalence class given in Eq.~(\ref{eq:ECrestriction}), so that
there are really only five degrees of freedom being fixed by the
choice of these variables.

Because all 2-metrics are conformally equivalent, specifying
$\FBMetric_{AB}$ can now clearly be seen as simply specifying the
coordinates on the excision surface.  The variables $\RadialLapse$ and
$\RadialShift^A$ then determine how these coordinates propagate off of
this boundary surface into the full initial-data hypersurface.  These
three degrees of freedom must be chosen so that the coordinates
satisfy the gauge conditions in Eq.~(\ref{eq:genconfharm}).  In terms
of our adapted coordinates, and making use of
Eqs.~(\ref{eq:ConfConns}) and (\ref{eq:FlatConns}), these three gauge
conditions are
\begin{eqnarray} 
\label{eq:Vr}
\GaugeFunc^r &=& -\frac1{\sqrt\FMetric}
     \partial_r\left(\sqrt\FMetric\RadialLapse^{-2}\right)
     + \FBCD_A\left(\RadialLapse^{-2}\RadialShift^A\right)
\\ \nonumber && \mbox{}
     + \RadialLapse\partial_r\ln\sqrt\FBMetric
     + \frac12\RadialLapse^{-2}\RadialShift^A\RadialShift^B
        \partial_r\FBMetric_{AB}, \\
\label{eq:VA}
\GaugeFunc^A &=& \frac1{\sqrt\FMetric}
     \partial_r\left(\sqrt\FMetric\RadialLapse^{-2}\RadialShift^A\right)
     - \FBCD_B\left(\RadialLapse\FBMetric^{AB} 
         + \RadialLapse^{-2}\RadialShift^A\RadialShift^B\right)
\nonumber \\ && \mbox{}
     + \RadialLapse^{-2}\RadialShift^B\FBMetric^{AC}\partial_r\FBMetric_{BC},
\end{eqnarray}
and again for compactness we define $\det(\FMetric)\equiv\FMetric$ and
$\det(\FBMetric)\equiv\FBMetric$.

As an explicit example, assume we are using Cartesian coordinates on
the spatial hypersurface.  For simplicity, consider the case where the
gauge source functions are chosen to vanish $(\GaugeFunc^i=0)$ and the
excision boundary is chosen to be a coordinate sphere of radius $r$.
Then the coordinates adapted to the excision surface are standard
spherical coordinates, we have
\begin{equation}
  S_{AB} \Rightarrow \left[\begin{array}{cc}
     r^2 & 0 \\
      0  & r^2\sin^2\theta \end{array}\right]
\end{equation} 
and $\det(f)=\det(S)\Rightarrow r^4\sin^2\theta$, and the coordinate
transformations between the three-dimensional Cartesian coordinates
and our adapted coordinates are well known.  The three gauge conditions
then take the form
\begin{eqnarray}
\label{eq:ExampleBCr}
  \CBnormal^i\partial_i\RadialLapse &=& \frac1r(1 - \RadialLapse^3) 
  - \frac12\left(\FBCD_A\RadialShift^A 
      + \frac1r\RadialShift_A\RadialShift^A\right) \\
\label{eq:ExampleBCA}
\CBnormal^i\partial_i\RadialShift^A &=& 
     \RadialLapse\FBCD^A\RadialLapse
     - \frac2{r\RadialLapse}\left(1 + \RadialLapse^3 
               + \frac12\RadialShift_B\RadialShift^B\right)\RadialShift^A
\\ \nonumber && \mbox{} \hspace{0.25in}
     + \frac1\RadialLapse\RadialShift^B\RadialShift^C\FBcon{A}_{BC}.
\end{eqnarray}
Note that these conditions are identically satisfied if the conformal
metric is flat so that $\RadialLapse=1$ and $\RadialShift^A=0$.

In this example, we impose boundary conditions on the six Cartesian
components of the conformal metric $\CMetric_{ij}$ as follows.  First
we use standard Cartesian-to-spherical coordinate transformations to
construct the three components of $\CMetric_{AB}=\CBMetric_{AB}$.  In
a similar fashion, we construct the two components of $\RadialShift^A$
and the scalar $\RadialLapse$.  The variables $\CBMetric_{AB}$,
$\RadialShift^A$, and $\RadialLapse$ are then explicit functions of
the six Cartesian components of the unknown conformal metric
$\CMetric_{ij}$.  We then demand that $\CBMetric_{AB} =
S_{AB}/\RadialLapse$ which constitute three Dirichlet boundary
conditions.  Finally, we impose the three normal-derivative boundary
conditions given in Eqs.~(\ref{eq:ExampleBCr}) and
(\ref{eq:ExampleBCA}).  Together, these form a coupled set of six
equations for the six components of $\CMetric_{ij}$.  However, we
emphasize that the choice of $\FBMetric_{AB}$ guarantees that the
global condition of $\det(\CMetric)$ is satisfied, and so we are
really only constraining the five remaining degrees of freedom in the
conformal metric.

Finally, we note that when implementing boundary conditions on the
Cartesian components of the conformal metric, one would not actually
use the boundary conditions as written in Eqs.~(\ref{eq:ExampleBCr})
and (\ref{eq:ExampleBCA}).  Instead, one would use the gauge
conditions of Eq.~(\ref{eq:genconfharm}) directly in terms of
Cartesian components.  The $2+1$ decomposition we have derived
is actually most useful in implementing the Dirichlet conditions,
and in better understanding the nature of the boundary conditions.

\section{Summary and Discussion}
\label{sec:summary}

In this paper we derive boundary conditions for the conformal metric
that can be applied on black-hole excision boundaries.  These boundary
conditions are needed in the context of a new initial-data
decomposition recently proposed by SUF, which provides an equation for
the conformal metric rather than treating it as a freely specifiable
variable.  This seems very attractive, since it avoids the need to make
ad-hoc choices for the conformal metric -- like conformal flatness --
and instead computes the conformal metric as part of the solution.

Whenever black-hole initial data are constructed using excision
methods, boundary conditions for all constrained data are required on
the excision boundaries.  In the conformal thin-sandwich
decomposition, boundary conditions for the conformal factor $\CF$ and
shift $\Shift^i$ had a strong physical motivation that can be
naturally viewed within the isolated-horizons
framework \cite{Cook-2002,cook-pfeiffer-2004a,Jaramillo-etal:2004}.
Interestingly, this is not the case for the boundary conditions
for the conformal metric. 

It might seem that different choices for the induced metric $S_{AB}$
should lead to physically different initial data.  While we have
emphasized the conformal equivalence of all closed 2-metrics, we must
recognize that this is a statement about the {\em intrinsic} geometry of the
2-surface only.  We must also consider the effect of the choice of the {\em
extrinsic} curvature.  Although we are restricted to choosing a
boundary 2-surface with topology $S^2$, we have a wide range of
choices for how this surface is embedded in the reference space.  For
example, we could choose the boundary to be either a coordinate sphere
or a coordinate ellipse in a flat Cartesian space,  each of which has a
different extrinsic curvature.  Note also that we can make this choice 
independently of the choice of the metric on the excision surface; 
for example, we could choose $S_{AB}$ to be the metric of a unit sphere
even on a coordinate ellipse.  Both the choice for the shape of
the 2-surface and for $S_{AB}$ will affect the {\em extrinsic}
curvature of the 2-surface as defined by Eq.~(\ref{eq:SurfExCurv}).
But, it is important to remember that these fix the {\em conformal}
extrinsic curvature of the 2-surface, and not its {\em physical} extrinsic
curvature.

The 2-surface's influence on the dynamical degrees of freedom of our
initial data are embodied in its physical extrinsic curvature via the
sheer and expansion of the family of outgoing null rays passing
through the boundary 2-surface.  The physically motivated boundary
conditions on the conformal factor $\CF$ and shift $\Shift^i$
mentioned above are, in fact, obtained by demanding the sheer and
expansion vanish on the boundary.  These choices will certainly have
an affect on the physical extrinsic curvature of the excision surface.
Furthermore, the choice of the shape of the excision surface may have
some affect on the dynamical degrees of freedom of the initial data,
although the choice for $S_{AB}$ should not.

As we demonstrate in this paper, it is natural to write the conformal
metric on the excision surface in a 2+1 decomposition.  The induced
conformal metric on the excision surface must be related by suitable
coordinate and conformal transformations to the metric on a unit
sphere -- making this choice therefore does not impose any physical
restriction on the conformal metric.  The remaining degrees of freedom
of the conformal metric on the excision surface can be expressed as a
radial ``lapse" and ``shift", and can be determined from the gauge
conditions that are imposed on the conformal metric.  We therefore
conclude that these boundary conditions affect only the gauge degrees
of freedom of the conformal metric and in no way restrict the
dynamical degrees of freedom of our initial-data solution.

While the investigation of these boundary conditions was motivated by
the need to provide excision boundary conditions for black-hole
initial data, the fact that only general geometrical and gauge
considerations are involved means that these boundary conditions can
be applied to any topologically spherical boundary.  These boundary
conditions need not be applied at a black-hole horizon, and in
particular these conditions can be applied without approximation on an
outer boundary.

Finally, we repeat our warning from Sec.~\ref{sec:introduction} that
the set of boundary conditions we have derived have not yet been shown
to be independent and give rise to a well-posed elliptic system.
Because the set of elliptic equations are non-linear and coupled, it
is very difficult (if even possible) to determine this analytically.
Future numerical implementations will clarify this issue.

\acknowledgments

The authors are grateful to J.\ Isenberg, L.\ Lindblom, V.\ Moncrief,
N.\ \`O Murchadha, M.\ Scheel, and M.\ Shibata for illuminating
discussions and are grateful to K.\ Thorne and L. Lindblom for their
hospitality during the Caltech Visitors Program in the Numerical
Simulation of Gravitational Wave Sources, during which work on this
idea began.  The authors gratefully acknowledge the hospitality at and
support from the Kavli Institute of Theoretical Physics at the
University of California Santa Barbara.  This work was supported in
part by NSF grants PHY-0555617 to Wake Forest University, and
PHY-0456917 and PHY-0756514 to Bowdoin College.  G.B.C. acknowledges
support from the Z.\ Smith Reynolds Foundation.

\begin{widetext}
\appendix
\section{Einstein's Equations in Reference Metric Form}
\label{sec:app}

For completeness, we include below the conformally decomposed
$3+1$ version of Einstein's equations, expressed in a reference metric form, along with a few additional
useful equations.  In writing these equations, we assume that the
fundamental independent variables are
\begin{equation}
  \CF,\quad\CLapse,\quad\Shift^i,\quad\text{and}\quad\CMetric^{ij},
\end{equation}
while we take the following quantities to be freely specifiable
\begin{equation}
  \dtime\CMetric^{ij},\quad\dtime\CTFExCurv^{ij},\quad\dtime\TrExCurv,
  \quad\TrExCurv,\quad\det\CMetric,\quad\text{and}\quad\GaugeFunc^i.
\end{equation}
The conformal metric $\CMetric_{ij}$ and the trace-free conformal
extrinsic curvature $\CTFExCurv^{ij}$ are auxiliary variables defined
to simplify the equations.  The conformal metric is derived from the
conformal inverse metric $\CMetric^{ij}$ in the usual way, and the
trace-free conformal extrinsic curvature is defined by
Eq.~(\ref{eq:ExCurvRMdef}) below.  Note that, as in the main text,
derivatives of the gauge source functions $\GaugeFunc^i$ are taken
using the conformal covariant derivative $\CCD_i$ whereas derivatives
of the main variables are taken using the flat covariant derivative
$\FCD_i$.  We emphasize again that we fix the reference metric to be
flat (although this can be generalized if needed) and therefore the
Ricci tensor and scalar associated with the reference metric vanish.

The Hamiltonian constraint in Eq.~(\ref{eq:CHamCon1}) is an
elliptic equation for the conformal factor and takes the form
\begin{equation}
\RLap\CF - \mbox{$\frac1{16}$}\CF\CMetric_{ij}
({\cal C}^{ij} - \mbox{$\frac12$}{\cal B}^{ij})
- \mbox{$\frac1{12}$}\CF^5\TrExCurv^2
+ \mbox{$\frac18$}\CF^{-7}\CMetric_{ij}\CMetric_{k\ell}
\CTFExCurv^{ik}\CTFExCurv^{j\ell} 
= \GaugeFunc^i\FCD_i\CF + \mbox{$\frac18$}\CF\CCD_i\GaugeFunc^i
- \mbox{$\frac1{16}$}\CF\RLap\ln\CMetric,
\end{equation}
and the momentum constraint in Eq.~(\ref{eq:CMomCon1}) becomes
\begin{equation}
\label{eq:MomConRM1}
\FCD_j\CTFExCurv^{ij} - \mbox{$\frac23$}\CF^6\CMetric^{ij}\FCD_j\TrExCurv
+ \mbox{$\frac12$}\CTFExCurv^{km}\CMetric_{k\ell}\CMetric_{mn}
\CMetric^{ij}\FCD_j\CMetric^{\ell n}
- \CTFExCurv^{k\ell}\CMetric_{jk}\FCD_\ell\CMetric^{ij}
= - \mbox{$\frac12$}\CTFExCurv^{ij}\FCD_j\ln\CMetric.
\end{equation}

The trace-free conformal extrinsic curvature is defined in terms of
the time derivative of the inverse-conformal metric
\begin{equation}
\label{eq:ExCurvRMdef}
\dtime\CMetric^{ij} = 2\CLapse\CTFExCurv^{ij}
- 2\CMetric^{k(i}\FCD_k\Shift^{j)}
+ \mbox{$\frac23$}\CMetric^{ij}\FCD_k\Shift^k
+ \Shift^k\FCD_k\CMetric^{ij}
+ \mbox{$\frac13$}\CMetric^{ij}\Shift^k\FCD_k\ln\CMetric 
\equiv -\CMtd^{ij}.
\end{equation}
Note the definition of $\CMtd^{ij}$ made for convenience below.  The
evolution equation for the trace-free conformal extrinsic curvature
Eq.~(\ref{eq:dtCA1}) becomes an elliptic equation for the inverse
conformal metric:
\begin{eqnarray}
\label{eq:dtExCurvRM}
\dtime\CTFExCurv^{ij} &=&
\mbox{$\frac12$}\CF^8\CLapse\left[\RLap\CMetric^{ij} - {\cal B}^{ij}
- {\cal C}^{ij} - \mbox{$\frac13$}\CMetric^{ij}\CMetric_{k\ell}{\cal C}^{k\ell}
+ 2{\cal E}^{ij}
- \mbox{$\frac12$}({\cal D}^{ij}
- \mbox{$\frac13$}\CMetric^{ij}\CMetric_{k\ell}{\cal D}^{k\ell})\right]
\\ \nonumber && \mbox{}
- (\CMetric^{ik}\CMetric^{j\ell} 
- \mbox{$\frac13$}\CMetric^{ij}\CMetric^{k\ell})\FCD_k\FCD_\ell(\CF^8\CLapse)
- (\CMetric^{\ell(i}\FCD_\ell\CMetric^{j)m}
- \mbox{$\frac12$}\CMetric^{\ell m}\FCD_\ell\CMetric^{ij})\FCD_m(\CF^8\CLapse)
\\ \nonumber && \mbox{}
+ 8\CF^8\CLapse(\CMetric^{k(i}\CMetric^{j)\ell} 
- \mbox{$\frac13$}\CMetric^{ij}\CMetric^{k\ell})(\FCD_k\ln\CF)
  \FCD_\ell\ln(\CF^7\CLapse)
\\ \nonumber && \mbox{}
+ \Shift^k\FCD_k\CTFExCurv^{ij} + \CTFExCurv^{ij}\FCD_k\Shift^k
+ \CMetric_{\ell m}\CTFExCurv^{m(i}\CMetric^{j)k}\FCD_k\Shift^\ell
- \CTFExCurv^{k(i}\FCD_k\Shift^{j)}
- \Shift^k\CMetric_{\ell m}\CTFExCurv^{\ell(i}\FCD_k\CMetric^{j)m}
- \CMetric_{k\ell}\CTFExCurv^{k(i}\CMtd^{j)\ell}
\\ \nonumber && \mbox{}
+ \mbox{$\frac12$}\CF^8\CLapse\CLD{\GaugeFunc}^{ij}
- \mbox{$\frac13$}\CMetric^{ij}\GaugeFunc^k\FCD_k(\CF^8\CLapse)
+ \mbox{$\frac16$}\CF^8\CLapse\CMetric^{ij}\RLap\ln\CMetric
+ \mbox{$\frac12$}\Shift^k\CTFExCurv^{ij}\FCD_k\ln\CMetric,
\end{eqnarray}
while the evolution equation for the trace of the extrinsic curvature
Eq.~(\ref{eq:dtTrK1}) is an elliptic equation for the conformal lapse
\begin{eqnarray}
\dtime\TrExCurv &=& - \CF^{-5}\RLap(\CF^7\CLapse) 
+ \mbox{$\frac1{16}$}\CF^2\CLapse\CMetric_{ij}
({\cal C}^{ij} - \mbox{$\frac12$}{\cal B}^{ij})
+ \mbox{$\frac5{12}$}\CF^6\CLapse\TrExCurv^2
+ \mbox{$\frac78$}\CF^{-6}\CLapse\CMetric_{ij}\CMetric_{k\ell}
\CTFExCurv^{ik}\CTFExCurv^{j\ell} 
+ \Shift^i\FCD_i\TrExCurv \\ \nonumber && \mbox{}
+ \CF^{-5}\GaugeFunc^i\FCD_i(\CF^7\CLapse) 
+ \mbox{$\frac18$}\CF^2\CLapse\CCD_i\GaugeFunc^i
- \mbox{$\frac1{16}$}\CF^2\CLapse\RLap\ln\CMetric.
\end{eqnarray}
Finally, the evolution equation for the conformal factor becomes
\begin{equation}
\dtime\CF = \mbox{$\frac16$}\CF\left(\FCD_i\Shift^i
+ 6\Shift^i\FCD_i\ln\CF - \CF^6\CLapse\TrExCurv\right)
+ \mbox{$\frac1{12}$}\CF\Shift^i\FCD_i\ln\CMetric.
\end{equation}

If we replace $\CTFExCurv^{ij}$ in the momentum constraint
(\ref{eq:MomConRM1}) by its definition in Eq.~(\ref{eq:ExCurvRMdef}),
we obtain an elliptic equation for the shift:
\begin{eqnarray}
\RLap\Shift^i + \mbox{$\frac13$}\CMetric^{ij}\FCD_j\FCD_k\Shift^k
+ \mbox{$\frac12$}(2\CLapse\CTFExCurv^{km})\CMetric_{k\ell}\CMetric_{mn}
\CMetric^{ij}\FCD_j\CMetric^{\ell n}
- (2\CLapse\CTFExCurv^{k\ell})\CMetric_{jk}\FCD_\ell\CMetric^{ij}
- (2\CLapse\CTFExCurv^{ij})\FCD_j\ln\CLapse
- \mbox{$\frac43$}\CLapse\CF^6\CMetric^{ij}\FCD_j\TrExCurv
- \FCD_j\CMtd^{ij}
\hspace{0.05in}\\ \nonumber
= \GaugeFunc^j\FCD_j\Shift^i - \mbox{$\frac23$}\GaugeFunc^i\FCD_j\Shift^j
- \Shift^j\FCD_j\GaugeFunc^i 
- \mbox{$\frac13$}\GaugeFunc^i\Shift^j\FCD_j\ln\CMetric
- \mbox{$\frac12$}(2\CLapse\CTFExCurv^{ij})\FCD_j\ln\CMetric
+ \mbox{$\frac13$}\CMetric^{ij}(\FCD_j\Shift^k)\FCD_k\ln\CMetric
- \mbox{$\frac13$}\CMetric^{ij}(\FCD_k\Shift^k)\FCD_j\ln\CMetric
\\ \nonumber 
+ \mbox{$\frac12$}\CMetric^{kj}(\FCD_k\Shift^i)\FCD_j\ln\CMetric
- \mbox{$\frac12$}\Shift^k(\FCD_k\CMetric^{ij})\FCD_j\ln\CMetric
- \mbox{$\frac16$}\CMetric^{ij}\Shift^k\FCD_j\FCD_k\ln\CMetric
- \mbox{$\frac16$}\CMetric^{ij}\Shift^k(\FCD_j\ln\CMetric)\FCD_k\ln\CMetric.
\end{eqnarray}

In all of these equations terms of the form $\FCD_i\ln\CMetric$ appear.
These terms are defined implicitly by
\begin{equation}
  \FCD_i\ln\CMetric \equiv \frac1\CMetric\FCD_i\CMetric
  = \partial_i\ln\left(\frac\CMetric{f}\right).
\end{equation}
Note that both $\det(\CMetric)$ and $\det(\FMetric)$ transform as
scalar densities of weight 2, so their ratio is a simple scalar.  If
we choose $\det(\CMetric)=\det(\FMetric)$ as we have in defining the
boundary boundary conditions in Sec.~\ref{sec:boundary-conditions},
then we find that all terms involving derivatives of $\det(\CMetric)$
vanish.

Finally, we note that the Arnowit-Deser-Misner energy takes on the
familiar form:
\begin{equation}
E_{ADM} = -\frac1{2\pi}\oint_\infty{\FCD_i\CF\,\td^2S^i}
+ \frac1{16\pi}\oint_\infty\left(\GaugeFunc^j 
- \mbox{$\frac12$}\CMetric^{ij}\FCD_i\ln\CMetric\right)\td^2S_j.
\end{equation}

\end{widetext}

\end{document}